# "Race and Gender"
## Timnit Gebru

**ABSTRACT**

From massive face-recognition-based surveillance and machine-learning-based decision systems predicting crime recidivism rates, to the move towards automated health diagnostic systems, artificial intelligence (AI) is being used in scenarios that have serious consequences in people's lives. However, this rapid permeation of AI into society has not been accompanied by a thorough investigation of the sociopolitical issues that cause certain groups of people to be harmed rather than advantaged by it. For instance, recent studies have shown that commercial face recognition systems have much higher error rates for dark skinned women while having minimal errors on light skinned men. A 2016 ProPublica investigation uncovered that machine learning based tools that assess crime recidivism rates in the US are biased against African Americans. Other studies show that natural language processing tools trained on newspapers exhibit societal biases (e.g. finishing the analogy "Man is to computer programmer as woman is to X" by *homemaker)*. At the same time, books such as *Weapons of Math Destruction* and *Automated Inequality* detail how people in lower socioeconomic classes in the US are subjected to more automated decision making tools than those who are in the upper class. Thus, these tools are most often used on people towards whom they exhibit the most bias. While many technical solutions have been proposed to alleviate bias in machine learning systems, we have to take a holistic and multifaceted approach. This includes standardization bodies determining what types of systems can be used in which scenarios, making sure that automated decision tools are created by people from diverse backgrounds, and understanding the historical and political factors that disadvantage certain groups who are subjected to these tools.

## 1. DATA-DRIVEN CLAIMS ABOUT RACE AND GENDER PERPETUATE THE NEGATIVE BIASES OF THE DAY

Science is often hailed as an objective discipline in pursuit of truth. Similarly, one may believe that technology is inherently neutral, and that products that are built by those representing only a slice of the world's population can be used by anyone in the world. However, an analysis of scientific thinking in the 19th century, and major technological advances such as automobiles, medical practices and other disciplines shows how the lack of representation among those who have the power to build this technology has resulted in a power imbalance in the world, and in technology whose intended or unintended negative consequences harm those who are not represented in its production[1]. Artificial intelligence is no different. While the popular paradigm of the day continues to change, the dominance of those who are the most powerful race/ethnicity in their location (e.g. White in the US, ethnic Han in China, etc.), combined with the concentration of power in a few locations around the world, has resulted in a technology that can benefit humanity but also has been shown to (intentionally or unintentionally) systematically discriminate against those who are already marginalized.

---

[1] O'Neil, Cathy. *Weapons of math destruction: How big data increases inequality and threatens democracy*. Broadway Books, 2016.

Like many disciplines, often those who perpetuate bias are doing it while attempting to come up with something better than before. However, the predominant thought that scientists are "objective" clouds them from being self-critical and analyzing what predominant discriminatory view of the day they could be encoding, or what goal they are helping advance. For example, in the 19th century, Charles Darwin worked on his theory of evolution as a carefully researched and well thought out alternative to creationism. What many leave out however is that the title of his book was *On the Origin of Species by Means of Natural Selection, or **the Preservation of Favoured Races in the Struggle for Life*** (emphasis added), in which he writes: "The western nations of Europe . . . now so immeasurably surpass their former savage progenitors [that they] stand at the summit of civilization. . . . [T]he civilised races of man will almost certainly exterminate, and replace, the savage races throughout the world."[2] And in his subsequent book, *The Descent of Man and Selection in Relation to Sex*, he notes that "[m]an is more courageous, pugnacious and energetic than woman, and has a more inventive genius. His brain is absolutely larger, [while] the formation of her skull is said to be intermediate between the child and the man."[3]

Although Darwin's book was criticized for its stance against the church, the British empire used it to justify colonialism by claiming that those subjected under its rule were scientifically inferior and unfit to rule themselves, with British anthropologists like James Hunt using Darwin's theory to justify slavery in papers such as *The Negro's Place in Nature* (1863).[4]

---

[2] Darwin, Charles. *On the Origin of Species by Means of Natural Selection, or the Preservation of Favoured Races in the Struggle for Life*. H. Milford; Oxford University Press, 1859
[3] Darwin, Charles. *The Descent of Man, and Selection in Relation to Sex*. Vol. 1. D. Appleton, 1896.
[4] Hunt, James. *On the Negro's Place in Nature*. Trübner, for the Anthropological Society, 1863.

Since the days of Darwin, race has been shown time and time again to be a social construct that has no biological basis.[5] According to professor of public health Michael Yudell, race is "a concept we think is too crude to provide useful information, it's a concept that has social meaning that interferes in the scientific understanding of human genetic diversity and it's a concept that we are not the first to call upon moving away from."[6]

However, celebrated scientists like evolutionary psychologist Steven Pinker still assert that it is tied to genetics, writing articles such as *Groups and Genes*[7] which claim, for example, that Ashkenazi Jews are innately intelligent. Echoing Darwin's assertions regarding the relationships between genius and gender, scientists are still attempting to extract gender based differences in intelligence with papers asking "Why are males over-represented at the upper extremes of intelligence?"[8]

These questions are posed without disputing the claim that males are overrepresented in the upper extremes of intelligence. Researchers have claimed to empirically show that men are overrepresented in the upper and lower extremes of IQ: that is, the highest and lowest scoring person in the IQ test is most likely to be a man.[9] This claim is then generalized to mean that men show a greater spread in "intelligence" generally, without constraining it to the IQ test.

Because of the myth of scientific objectivity, these types of claims that seem to be backed up by data and "science" are less likely to be scrutinized. Just like Darwin and Hunt, many scientists today perpetuate the view that there is an inherent difference between the abilities of various races and sexes. However, because their works seem to be corroborated by data and empirical experiments, these views are likely to gain credibility. What is not captured in any of these analyses is, for example, that the IQ test in and of itself was designed by White men whose concept of "smartness" or "genius" was shaped, centered and evaluated on specific types of White men.

In fact, standardized testing in general has a racist history in the United States, and Ben Hutchinson and Margaret Mitchell's *50 Years of Unfairness* discusses bodies of work from the civil rights movement era that were devoted to fairness in standardized testing[10]. The debates and proposals put forth at that time foreshadow those advanced within the AI ethics and fairness community today.

Thus, the types of data-driven claims about race and gender made by the likes of Darwin are still alive today, and will probably be for the foreseeable future. The only difference will be the method of choice used to "corroborate" such claims. In 2019, Reuters reported that Amazon shut down its automated hiring tool because it was found to be negatively biased against women[11]. According to Reuters, the tool "penalized resumes that included the word 'women's,' as in 'women's chess club captain.'" And it downgraded graduates of two all-women's colleges.

---

[10] Hutchinson, Ben, and Margaret Mitchell. "50 Years of Test (Un)fairness: Lessons for Machine Learning." In *Proceedings of the Conference on Fairness, Accountability, and Transparency*, pp. 49-58. ACM, 2019

[11] Dastin, Jeffrey. "Amazon scraps secret AI recruiting tool that showed bias against women." Reuters (2018). https://www.reuters.com/article/us-amazon-com-jobs-automation-insight/amazon-scraps-secret-ai-recruiting-tool-that-showed-bias-against-women-idUSKCN1MK08G

Analyzed within the context of the society it was built in, it is unsurprising that an automated hiring tool such as Amazon's would exhibit these types of biases. In 2018, workers at Google staged a walkout protesting the company's handling of sexual harassment. And shortly after, in 2019, news articles detailed women's accounts of toxic working environments at Microsoft including sexual harassment that goes unpunished, inability to get promoted, and many other forms of discrimination[12].

This hostile environment for women is ironic given the fact that the computing industry was started and dominated by women. As Marie Hicks details in *Programmed Inequality*, while computing was considered a feminine job dominated by women, that changed with the advent of the personal computer in the 1960s and 70s when computing started to be lucrative.[13]

This phenomenon is not unique to computing. Professions originally deemed by many societies to reflect women's tasks (e.g. cooking), cease to be regarded in this way when the work becomes lucrative. For example, the US restaurant business is dominated by men while cooking at home is still considered to be a woman's responsibility. Similarly, by the 1970s computing had gone from being considered a woman's job, to, within 20 years, one dominated by men. To select people who have innate "traits" of the successful programmer, IBM invented the Programmer Aptitude Test (PAT) which is similar to the IQ test[14]. Nathan Ensmenger notes that "[t]he focus on mathematical trivia, logic puzzles, and word games, for example, did not allow for any more

---

nuanced or meaningful or context-specific problem solving."[15] Sadly, until very recently, part of some companies' interview processes also involved solving these types of puzzles which have no connection to the job sought by the applicant. While some companies such as Google have eliminated the brain teasers after their own internal studies showed that they were not connected to the applicant's future success, many in the tech industry have adopted Google's style of whiteboard interviewing.

## 2. USING PAST DATA TO DETERMINE FUTURE OUTCOMES RESULTS IN RUNAWAY FEEDBACK LOOPS

An aptitude test designed by specific people is bound to inject their subjective biases of who is supposed to be good for the job, and eliminate diverse groups of people who do not fit the rigid, arbitrarily defined criteria that have been put in place. Those for whom the tech industry is known to be hostile will have difficulty succeeding, getting credit for their work, or promoted, which in turn can seem to corroborate the notion that they are not good at their jobs in the first place. It is thus unsurprising that in 2018, automated hiring tools used by Amazon and others which naively train models based on past data in order to determine future outcomes, create runaway feedback loops exacerbating existing societal biases.

A hiring model attempting to predict the characteristics determining a candidate's likelihood of success at Amazon would invariably learn that the undersampled majority (a term coined by Joy Buolamwini) are unlikely to succeed because the environment is known to be hostile towards people of African, Latinx, and Native American descent, women, those with disabilities, members of the LGBTQ+ community and any community that has been marginalized in the tech industry and in the US. The person may not be hired because of bias in the interview process, or may not succeed because of an environment that does not set up people from certain groups for success. Once a model is trained on this type of data, it exacerbates existing societal issues driving further marginalization.

---

[15] Ensmenger, Nathan L. *The computer boys take over: Computers, programmers, and the politics of technical expertise*. MIT Press, 2012.

The model selects for those in the non-marginalized group, who then have a better chance of getting hired because of a process that favors them, and a higher chance of success in the company because of an environment that benefits them. This generates more biased training data for the hiring tool which further reinforces the bias creating a runaway feedback loop of increasing the existing marginalization.

These types of feedback loops amplifying bias are not unique to hiring models. Predictive policing, predicting crime "hotspots" based on a model trained on data of who has been arrested in which neighborhood, or which crimes have been reported, has also been shown to exhibit runway feedback loops. In many parts of the US, there is a large discrepancy between who commits a crime vs. whose crimes are reported. For example, the national survey on drug use and health shows drug use to be relatively evenly spread out in Oakland, whereas reports of drug use to police are concentrated in predominantly Black neighborhoods. Kristian Lum and William Isaacs have shown that the popular predictive policing model, Predpol, reinforces existing inequities by predicting these predominantly Black neighborhoods to be crime hotspots[16]. More police are then sent to these neighborhoods, in which case they arrest more people from those locations than places with less police presence--seeming to validate the presence of more crime in those neighborhoods than others. These new arrests are then used as additional training data, increasing over-policing in disadvantaged neighborhoods and amplifying societal bias.

---

[16] Lum, Kristian, and William Isaac. "To predict and serve?." *Significance* 13, no. 5 (2016): 14-19.

**3. UNREGULATED USAGE OF BIASED AUTOMATED FACIAL ANALYSIS TOOLS**

Predictive policing is only one of the data-driven algorithms employed by US law enforcement. The perpetual lineup report by Clare Garvie, Alvaro Bedoya and Jonathan Frankle discusses law enforcement's unregulated use of face recognition in the United States, stating that one in two American adults are in a law enforcement database that can be searched and used at any time[17]. There is currently no regulation in place auditing the accuracy of these systems, or specifying how and when they can be used. The report further discusses the potential for people to be sent to jail due to cases of mistaken identity, and notes that operators are not well trained on using any of these tools. The authors propose a model law guiding government usage of automated facial analysis tools, describing a process by which the public can debate its pros and cons before it can be used by law enforcement.

As it stands, unregulated usage of automated facial analysis tools is spreading from law enforcement to other high stakes sectors such as employment. And a recent study by Buolamwini and Gebru shows that these tools could have systematic biases by skin-type and gender[18]. After analyzing the performance of commercial gender classification systems from 3 companies, Microsoft, Face++ and IBM, the study found near perfect classification for lighter skinned men (error rates of 0% to 0.8%), whereas error rates for darker skinned women were as high as 35.5%. After this study was published, Microsoft and IBM released new versions of their APIs less than 6 months after the paper's publication, major companies such as Google established fairness organizations, and US Senators Kamala Harris, Cory Booker and Cedric Richmond called for regulation on law enforcement's use of automated facial analysis tools[19]. Even those in the healthcare industry cautioned against the blind use of unregulated AI.

---

[17] Clare Garvie, Alvaro Bedoya, and Jonathan Frankle. The Perpetual Line-Up: Unregulated Police Face Recognition in America. Georgetown Law, Center on Privacy & Technology, 2016.
[18] Buolamwini, Joy, and Timnit Gebru. "Gender shades: Intersectional accuracy disparities in commercial gender classification." In *Conference on fairness, accountability and transparency*, pp. 77-91. 2018.

As shown in Buolamwini and Gebru's study, society's concept of race and gender affects the design and usage of AI systems. For example, although works prior to Gender shades: Intersectional accuracy disparities in commercial gender classification[18] have studied the accuracy of automated facial analysis tools by using geography as a proxy for race, none had performed the analysis by skin type, and none intersectionally—
taking into account multiple identities such as gender and skin type. As a duo of darker and lighter skinned Black women in the US, Buolamwini and Gebru understood that race is an unstable social construct across time and space, having different meanings in different cultures, locations and historical periods.

In *The Cost of Color*, sociologist Ellis Monk notes that "some studies even suggest that within-race inequalities associated with skin tone among African Americans often rival or exceed what obtains between blacks and whites as a whole."[20] Thus, instead of performing their analysis by race, Buolamwini and Gebru used the Fitzpatrick skin type classification system to classify images into darker and lighter skinned subjects, analyzing the accuracy of commercial systems for each of these subgroups.

---

[19] Harris D., Kamala, Cory A. Booker, and Cedric L. Richmond. Letter to the Federal Bureau of investigation. https://www.scribd.com/embeds/388920671/content#from_embed 2018.
[20] Monk Jr, E.P., 2015. The cost of color: Skin color, discrimination, and health among African-Americans. *American Journal of Sociology. 121*(2), pp 396-444.

Buolamwini and Gebru's work notes that AI systems need to be tested intersectionally to uncover their shortcomings. Kimberlé Crenshaw, a leading scholar who coined the term *intersectionality* in critical race theory, stresses the importance of taking into account an individual's different identities and how they interact with systems of power in tandem[21]. She often gives the example of a 1976 lawsuit by Emma DeGraffenreid alleging that General Motors (GM) discriminated against Black women. The plaintiffs lost the lawsuit with judges reasoning that since GM hires Black people, and also hires women, they couldn't have discriminated against Black women.

What they failed to see however is that GM hired women for secretarial positions but it wouldn't hire Black people for these positions. And GM hired men for factory positions, but didn't consider women for these positions. Thus, Black women were indeed discriminated against by GM, but without an intersectional view of both race *and* gender, the judges were unable to see this discrimination. In Buolamwini and Gebru's work, analyzing these systems by both gender and skin type showed the largest disparities and both women discuss their life experiences and understanding of works on intersectionality as their motivation for disaggregating accuracy by gender and skin type.

## 4. AI BASED TOOLS ARE PERPETUATING GENDER STEREOTYPES

---

[21] Crenshaw, Kimberle. "Demarginalizing the intersection of race and sex: A black feminist critique of antidiscrimination doctrine, feminist theory and antiracist politics." *U. Chi. Legal F.* (1989): 139

While the previous section has discussed manners in which automated facial analysis tools with unequal performance across different subgroups are being used by law enforcement, this section shows that the existence of some tools in the first place, no matter how "accurate" they are, can perpetuate harmful gender stereotypes.

There are many ways in which society's views of race and gender are encoded into the AI systems that are built. Studies such as Hamidi et al.'s *Gender Recognition or Gender Reductionism*[22] discuss this in the context of automatic gender recognition systems such as those studied by Buolamwini and Gebru, and the harms they cause particularly to the transgender community.

For instance, the task of automatic gender recognition (AGR) itself implicitly assumes that gender is a static concept that does not frequently change across time and cultures. However, gender presentations greatly differ across cultures--a fact that is often unaccounted for in these systems. Gender classification systems are often trained with data that has very few or no transgender and non-binary individuals. And the outputs themselves only classify images as "male" or "female." For transgender communities, the effects of AGR can be severe, ranging from misgendering an individual to outing them in public. Hamidi et al. note that according to the National Transgender Discrimination Survey conducted in 2014, 56% of the respondents who were regularly misgendered in the workplace had attempted suicide.[23] While there are well documented harms due to systems that perform AGR, the utility of these tools is often unclear.

---

[22] Hamidi, Foad, Morgan Klaus Scheuerman, and Stacy M. Branham. "Gender recognition or gender reductionism?: The social implications of embedded gender recognition systems." In *Proceedings of the 2018 CHI Conference on Human Factors in Computing Systems*, p. 8. ACM, 2018.

One of the most common applications of AGR is for targeted advertising (e.g. showing those perceived to be women a specific product). This has the danger of perpetuating stereotypes by giving subliminal messages regarding artifacts that men vs. women should use. For example, Urban Outfitters started personalizing their website based on the perceived genders of their frequent customers. But the program was scrapped after many customers objected to gender-based marketing: some shoppers often bought clothes that were not placed in their ascribed gender's section, and others were opposed to the concept of gender-based targeting in and of itself[24].

Automatic gender recognition systems are only one of the many ways in which stereotypes and gender based societal biases are propagated through AI. From the imagery used to visualize cyborgs, to the names, voices and mannerisms depicted by speech recognition systems like Siri and Alexa who are meant to obey a customer's every whim, it is clear that the design of commercial AI systems is based on stereotypical gender roles. Amy Chambers writes:

> Virtual assistants are increasingly popular and present in our everyday lives: literally with Alexa, Cortana, Holly, and Siri, and fictionally in films Samantha (Her), Joi (Blade Runner 2049) and Marvel's AIs, FRIDAY (Avengers: Infinity War), and Karen (Spider-Man: Homecoming). These names demonstrate the assumption that virtual assistants, from SatNav to Siri, will be voiced by a woman. This reinforces gender stereotypes, expectations, and assumptions about the future of artificial intelligence.[25]

---

[23] Hamidi, Foad, Morgan Klaus Scheuerman, and Stacy M. Branham. "Gender recognition or gender reductionism?: The social implications of embedded gender recognition systems." In *Proceedings of the 2018 CHI Conference on Human Factors in Computing Systems*, p. 8. ACM, 2018.

[24] Singer, Natasha. "E-tailer customization: convenient or creepy." *New York Times*, June 12, 2012.
[25] Chambers, Amy. "There's a reason Siri, Alexa and AI are imagined as female – sexism." *The Conversation* (2018). http://theconversation.com/theres-a-reason-siri-alexa-and-ai-are-imagined-as-female-sexism-96430

What does it mean for children to grow up in households filled with feminized voices that are in clearly subservient roles? AI systems are already used in ways that are demeaning to women without explicitly encoding gendered names and voices. For example, generative adversarial networks (GANs), models that have been used to generate imagery among many other things, have been weaponized against women[26]. Deep fakes, videos generated using GANs, create pornographic content using the faces of ordinary women whose photos have been scraped from social media without consent.

## 5. POWER IMBALANCE AND THE EXCLUSION OF MARGINALIZED VOICES IN AI

The weaponization of technology against certain groups, as well as its usage to maintain the status quo while being touted as a liberator of those without power, is not new to AI. In Model Cards for Model Reporting, Mitchell et al. note parallels to other industries where products were designed for a homogenous group of people[27]. From automobiles crash tested on dummies with prototypical adult "male" characteristics resulting in accidents that disproportionately killed women and children, to clinical trials that excluded many groups of people resulting in drugs that do not work or disproportionately negatively affect women, products that are built and tested on a homogenous group of people work best for that group. A 2018 *Newsweek* article highlighting scientist Charles Rotimi notes: "By 2009, fewer than 1 percent of the several hundred genome investigations included Africans," even though "African genomes are the most diverse of any on the planet."[28] Excluding African genes not only hurts those of African descent by creating next generation personalized drugs that do not work for them, but also leads scientists to erroneous claims by overfitting on homogenous data, by e.g. reaching conclusions based on uncommon mutations among European genomes but ones that are common in Africans.

---

[26] Curtis, Cara. "Deepfakes are being weaponized to silence women—but this woman is fighting back." *The Next Web* (2018). https://thenextweb.com/code-word/2018/10/05/deepfakes-are-being-weaponized-to-silence-women-but-this-woman-is-fighting-back/
[27] Mitchell, Margaret, Simone Wu, Andrew Zaldivar, Parker Barnes, Lucy Vasserman, Ben Hutchinson, Elena Spitzer, Inioluwa Deborah Raji, and Timnit Gebru. "Model cards for model reporting." In Proceedings of the Conference on Fairness, Accountability, and Transparency, pp. 220-229. ACM, 2019.

Indeed, the development and trajectory of AI seems to be mirroring many other disciplines. In a Blogpost, Ali Alkhatib describes the harm current AI development has caused to marginalized groups, and its parallels to anthropology.[29] He points out that "anthropologists, like computer scientists today, had the attention of the government - and specifically the military - and were drowning in lucrative funding arrangements. We were asked to do something that seemed reasonable at the time." Alkhatib cautions that "*the danger of aligning our work with existing power is the further subjugation and marginalization of the communities we ostensibly seek to understand*" (emphasis added), noting that "[t]he voices, opinions, and needs of disempowered stakeholders are being ignored today in favor of stakeholders with power, money, and influence—as they have been historically."

After a group of people from marginalized communities sacrificed their careers to shed light on how AI can negatively impact their communities, their ideas are now getting co-opted very quickly in what some have called a capture and neutralize strategy. In 2018 and 2019 respectively, The Massachusetts Institute of Technology (MIT) and Stanford University announced interdisciplinary initiatives centered around AI ethics, with multi-billion dollar funding from venture capitalists and other industries, and war criminals like Henry Kissinger taking center stage in both the Stanford and MIT opening events.

---

[28] Jessica, Wapner. "Cancer scientists have ignored African DNA in the search for cures." *Newsweek*, July 18 (2018).

[29] Alkhatib, Ali. "Anthropological/Artificial Intelligence & the HAI." (2019) https://ali-alkhatib.com/blog/anthropological-intelligence.

Mirroring what transpired in political anthropology, these well-funded initiatives exclude the voices of the marginalized people who they claim to support, and instead center powerful entities who have not worked on AI ethics, and in many cases have interests in proliferating unethical uses of AI. Like diversity and inclusion, ethics has become the language du jour. While Stanford's Human centered AI initiative has a mission statement that "[t]he creators of AI have to represent the world," the initiative was announced with zero Black faculty initially listed on the website out of 121 professors from multiple disciplines.

Universities are not the only institutions aspiring to be the central, authoritative voice on AI. Companies such as Amazon have announced a joint grant with The National Science Foundation (NSF) to fund fairness related research, while selling automated facial analysis tools with potentially systematic biases to law enforcement[30]. Shortly before the company announced its joint grant with the NSF, Amazon's leadership wrote a series of blog posts attempting to discredit the work of two Black women showing bias in their automated facial analysis tool[31].

While refusing to stop selling automated facial analysis tools to law enforcement without any regulation in place, and actively harming the careers of two women from marginalized communities negatively impacted by Amazon's product, the company then claims to work on fairness by announcing a joint grant with NSF. This incident is a microcosm for the capture and neutralize strategy that disempowers those from marginalized communities while using the fashionable language of ethics, fairness, diversity and inclusion to advance the needs of the corporation at all costs.

---

[30] Singer, Natasha. "Amazon Is Pushing Facial Technology That a Study Says Could Be Biased." *New York Times*, Jan. 24, 2019.
[31] Raji, Inioluwa Deborah, and Joy Buolamwini. "Actionable auditing: Investigating the impact of publicly naming biased performance results of commercial ai products." In AAAI/ACM Conf. on AI Ethics and Society, vol. 1. 2019.

A letter signed by 78 scientists[32] including 2019 Turing award winner Yoshua Bengio later detailed the misrepresentations by Amazon officials, stressing the importance of the study and calling on Amazon to cease selling Rekognition to law enforcement. It was initially written by Timnit Gebru and Margaret Mitchell, the former being a Black woman and a collaborator of Buolamwini and Raji. This activism shows a bifurcation between the people who are taking risks within the work of ethics and fairness, vs. those who are given a seat at the table and centered in initiatives like MIT and Stanford. While two Black women pointed out the systematic issues with Amazon's products, and a third assembled a coalition of AI experts to reinforce their message, many in the academic community continue to publish papers, and do research on AI and ethics in the abstract. As of 2019, fairness and ethics have become safe-to-use buzz words, with many in the machine learning community describing them as "hot" topic areas. However, few people working in the field question whether some technologies should exist in the first place, and often do not center the voices of those most impacted by the technologies they claim to make more "fair". For example, at least seven out of the nine organizers on a 2018 workshop on the topic of ethical, social and governance issues in AI[33] at a leading machine learning conference, Neural Information Processing Systems (NeurIPS), were White. If an entire field of research uses the pain of negatively impacted communities, co-opts their framework for describing their struggle, and uses it for the career advancement of those from communities with power, the field contributes to the further marginalization of communities rather than helping them. The current movement towards sidelining many groups in favor of powerful interests who have never thought about AI ethics except in the abstract, or have only been forced to confront it because of works from people in marginalized communities like Raji and Buolamwini, shows that the fairness, transparency, accountability and ethics in AI movement is on the road to doing "parachute science" like many of the fields before it. Ali Alkhatib writes:

> Computer scientists have utterly failed to learn from the history of other fields, and in doing so we're replicating the same morally objectionable, deeply problematic relationships that other fields could have warned us to avoid - indeed, have tried to warn others to avoid. Political anthropologists of the 1940s "tended to take colonial domination itself for granted", and in doing so fashioned itself principally as a tool to further that hegemonic influence by finding ways to shape indigenous cultures to colonial powers.[34]

---

[32] Concerned Researchers. "On Recent Research Auditing Commercial Facial Analysis Technology", *Medium* (2019). https://link.medium.com/REW0dWzNAY
[33] Workshop on Ethical, Social and Governance Issues in AI, *NeurIPS* (2018).

This colonial attitude is currently pervasive in the AI ethics space. Some have coined the terms "parachute research" or "helicopter research"[35] to describe scientists who "parachute in" to different marginalized communities, take what they would like for their work whether it is data, surveys or specimens, and leave. This type of work not only results in subpar science due to researchers who conduct it without understanding the context, but it further marginalizes the communities by treating them as caged curiosities (as mentioned by Joy Buolamwini) without alleviating their pain. The best way to help a community is by elevating the voices of those who are working to make their community better—not by doing parachute research. Academics who are serious about AI ethics, thus, need to ensure that they center the voices of those who they write about in the introduction paragraphs and motivation sections of their research papers. They should work to create space for those who are marginalized and amplify their voices, rather than using them to advance one's own career, and raise money from venture capitalists in their name.

---

[34] Alkhatib, Ali. "Anthropological/Artificial Intelligence & the HAI." (2019) https://ali-alkhatib.com/blog/anthropological-intelligence (citation omitted).
[35] Campbell, Theresa Diane. "A clash of paradigms? Western and indigenous views on health research involving Aboriginal peoples." *Nurse Researcher* 21, no. 6 (2014)

## 6. THE DESIGN OF ETHICAL AI STARTS FROM WHO IS GIVEN A SEAT AT THE TABLE

Ethical AI is not an abstract concept, but one that is in dire need of a holistic approach. It starts from who is at the table, who is creating the technology, and who is framing the goals and values of AI. As such, an approach that is solely crafted, led, and evangelized by those in powerful positions around the world, is bound to fail. Who creates the technology determines whose values are embedded in it.

For instance, if the tech industry were not dominated by cis gendered straight men, would we have developed automatic gender recognition tools that have been shown to harm transgender communities and encourage stereotypical gender roles? If they were the ones overrepresented in the development of artificial intelligence, what types of tools would we have developed instead? If the most significant input for developing AI used in the criminal justice system came from those who were wrongfully accused of a crime and confronted with high cash bail due to risk assessment scores, would we have had the algorithms of today that disproportionately disenfranchise Black and Brown communities in the US? If the majority of AI research were funded by government agencies working on healthcare rather than military entities such as the Defense Advanced Research Projects Agency (DARPA), would we be working towards drones that identify persons of interest?

A recent example of a Palestinian arrested for writing "good morning" in Arabic which was translated to "hurt them" in English or "attack them" in Hebrew by Facebook Translate shows some of the structural issues at play[36]. The person was arrested by Israeli authorities who later released him after verifying that he had indeed written "good morning." According to *Ha'aretz*, no one had checked the original Arabic version before arresting the individual. There are many issues that led to these series of events.

To start, had the field of language translation been dominated by Palestinians as well as those from other Arabic speaking populations, it is difficult to imagine that this type of mistake in the translation system would have transpired. Tools used by Google and Facebook currently work best for translations between English and other western languages such as French, reflecting which cultures are most represented within the machine learning and natural language processing communities. Most of the papers and corpora published in this domain focus on languages that are deemed important by those in the research community, those who have funding and resources, and companies such as Facebook and Google which are located in Silicon Valley. It is thus not surprising that the overwhelming bias of the researchers and the community itself is towards solving translation problems between languages such as French and English.

---

[36] Berger, Yotam. "Israel arrests Palestinian because Facebook translated 'good morning' to 'attack them'." *Ha'aretz,* Oct. 22, 2017.

Secondly, natural language processing tools embed the societal biases encoded in the data they are trained on. While Arab speaking people are stereotyped as terrorists in many non-Arab majority countries to the point that a math professor was interrogated on a flight due to a neighboring passenger mistaking his math writings for Arabic[37], similar stereotypes do not exist with the majority of English, French or other western language speakers. Thus, even when mistakes occur in translations between languages such as French and English, they are unlikely to have such negative connotations as mistaking "good morning" for "attack them."

Racial and gender biases in natural language processing tools are well documented. As shown by Bolukbasi et al. and Caliskan et al., word embeddings that were trained on corpora such as newspaper articles or books exhibit behaviors that are in line with the societal biases encoded by the training data[38]. For example, Bolukbasi et al. found that word embeddings could be used to generate analogies, and those trained on Google news complete the sentence "man is to computer programmer as woman is to "X" with "homemaker."[39] Similarly, Caliskan et al. demonstrated that in word embeddings trained from crawling the web, African American names are more associated with unpleasant concepts like sickness whereas European American names are associated with pleasant concepts like flowers.[40] Dixon et al.[41] have also shown that sentiment analysis tools often classify texts pertaining to LGBTQ+ individuals as negative. Given the stereotyping of Muslims as terrorists by many western nations, it is thus less surprising to have a mistake resulting in a translation to "attack them". This incident also highlights automation bias: the tendency of people to over-trust automated tools. An experiment designed by scientists at Georgia Tech University to examine the extent to which participants trust a robot, showed that they were willing to follow it towards what seemed to be a burning building, using pathways that were clearly inconvenient[42]. In the case of the Palestinian who was arrested for his "good morning" post, authorities *trusted* the translation system and did not think to first see the original text before arresting the individual.

---

[37] Staff, Guardian. "Professor: Flight was delayed because my equations raised terror fears." *The Guardian*, May 7, 2016.
[38] Bolukbasi, Tolga, Kai-Wei Chang, James Y. Zou, Venkatesh Saligrama, and Adam T. Kalai. "Man is to computer programmer as woman is to homemaker? debiasing word embeddings." In *Advances in neural information processing systems*, pp. 4349-4357. 2016; Islam, Aylin Caliskan, Joanna J. Bryson, and Arvind Narayanan. "Semantics derived automatically from language corpora necessarily contain human biases." *CoRR, abs/1608.07187* (2016).
[39] Bolukbasi, Tolga, Kai-Wei Chang, James Y. Zou, Venkatesh Saligrama, and Adam T. Kalai. "Man is to computer programmer as woman is to homemaker? debiasing word embeddings." In *Advances in neural information processing systems*, pp. 4349-4357. 2016.

One cannot ignore the structural issues at play while analyzing what happened here. In addition to the increased likelihood of errors in translating Palestinian Arabic dialects, the oppression of Palestinians also makes it more likely that whatever translation errors that do exist are more harmful towards them. Similar to the Google Photos incident that classified a Black couple as "gorillas", this translation system was most harmful because of the type of error it made.

---

[40] Islam, Aylin Caliskan, Joanna J. Bryson, and Arvind Narayanan. "Semantics derived automatically from language corpora necessarily contain human biases." *CoRR, abs/1608.07187* (2016).
[41] Dixon, Lucas, John Li, Jeffrey Sorensen, Nithum Thain, and Lucy Vasserman. "Measuring and mitigating unintended bias in text classification." In *Proceedings of the 2018 AAAI/ACM Conference on AI, Ethics, and Society*, pp. 67-73. ACM, 2
[42] Robinette, Paul, Wenchen Li, Robert Allen, Ayanna M. Howard, and Alan R. Wagner. "Overtrust of robots in emergency evacuation scenarios." In *The Eleventh ACM/IEEE International Conference on Human Robot Interaction*, pp. 101-108. IEEE Press, 2016.

In the Google Photos incident, there were as many instances of white people being mistaken for whales as Black people being misclassified as gorillas. However, the connotation of being mistaken for a whale is not rooted in racist and discriminatory history such as Black people being depicted as monkeys and gorillas[43]. Even if someone could convince themselves that algorithms sometimes just spit out nonsense, the structure of the nonsense will tend vaguely toward the structure of historical prejudices.

The dominance of certain groups and underrepresentation of others in natural language processing, computer vision and machine learning ensures that the problems these groups work on do not address the biggest challenges faced by those who are not part of the dominant group in the field. In fact, it can contribute to the further marginalization of these groups. The error of "good morning" being translated to "attack them" would not have had such grave consequences had the structural imbalance in power not made it such that a Palestinian was more likely to be surveilled and subjected to automated tools. Similarly, Black people and other marginalized communities in the United States are more likely to be subjected to surveillance and interact with automated tools than other groups[44]. And the systematic errors encoding bias and stereotypes (due to the datasets that are used and the demographic makeup of researchers and practitioners in this area), can be much more costly for those in marginalized communities than other groups. The existing power imbalance coupled with these types of systematic errors disproportionately affecting marginalized groups, makes proposals such as the extreme vetting initiative by the

---

[43] Hund, Wulf D., Charles W. Mills, and Silvia Sebastiani, eds. *Simianization: Apes, Gender, Class, and Race*. Vol. 6. LIT Verlag Münster, 2015.
[44] Eubanks, Virginia. *Automating inequality: How high-tech tools profile, police, and punish the poor*. St. Martin's Press, 2018.

United States Immigration and Customs Enforcement (ICE) even more problematic and scary. The 2018 initiative proposes that ICE partners with tech companies to monitor various people's social network data with automated tools, and use that analysis to decide whether they should be allowed to immigrate to the US, are expected to be good citizens, or are considered to be at risk of becoming terrorists. While any attempt to predict a person's future criminal actions is a dangerous direction to move towards warned by science fiction movies such as *Minority Report* and TV series like *Black Mirror*, the proposal is even scarier paired with the systematic errors of the automated tools that would be used for such analyses. Natural language processing and computer vision based tools have disproportionate errors and biases towards those who are already marginalized and are likely to be targeted by agencies such as ICE.

It is heartening to see that a group of 54 leading scientists in AI wrote a letter against the extreme vetting initiative[45]. However, the initiative has continued and only a few groups of people within the AI community, those who are developing the tools used in these practices, are truly speaking out against proposals such as this one. The extreme underrepresentation of marginalized groups in the latter community makes it even more difficult for them to care. And those who do speak up are from groups who are already facing a disproportionate amount of the burden to diversify and educate their own communities—adding to the minority tax that they already face.

## 7. EDUCATION IN SCIENCE AND ENGINEERING NEEDS TO MOVE AWAY FROM "THE VIEW FROM NOWHERE"

---

[45] Technology Experts Letter to DHS Opposing the Extreme Vetting Initiative, 2017.

If we are to work on technology that is beneficial to all of society, it has to start from the involvement of people from many walks of life and geographic locations. The future of who technology benefits will depend on who builds it and who utilizes it. As we have seen, the gendered and racialized values of the society in which this technology has been largely developed have seeped into many aspects of its characteristics. To work on steering AI in the right direction, scientists must understand that their science cannot be divorced from the world's geopolitical landscape, and there are no such things as meritocracy and objectivity. Feminists have long critiqued "the view from nowhere": the belief that science is about finding objective "truths" without taking people's lived experiences into account. This and the myth of meritocracy are the dominant paradigms followed by disciplines pertaining to science and technology that continue to be dominated by men. In *Replacing the "View from Nowhere,"* Sarah Marie Stitzlein writes:

> According to most feminists and some pragmatists, the acknowledgment of both subject and object as historically and politically situated requires that the subjects and objects of knowledge be placed on a more level playing field. When this is done, objectivity, as a form of responding to the rights and well being of fellow subjects as well as the objects of scientific inquiry, must be considered. Objectivity, then, is achieved to the extent that responsibility in inquiry is fulfilled and expanded. It follows that scientists must be held accountable for the results of their projects and that scientists must acknowledge the political nature of their work. Objectivity understood as such implies relationships between people, objects, and inquiry projects as central to its conception.[46]

The educational system must move away from the total abstraction of science and technology and instead show how people's lived experiences have contributed to the trajectory that technology follows. In his paper *The Moral Character of Cryptographic Work*, Phillip

---

[46] Stitzlein, Sarah M. "Replacing the 'View from Nowhere': A Pragmatist-Feminist Science Classroom." Electronic Journal of Science Education 9, no. 2 (2004) (citations omitted).

Rogaway sees the rise of mass surveillance as a failure of the cryptographic community[47]. He discusses various methods proposed in cryptography, and outlines how the extreme abstraction of the field and lack of accounting for the geopolitical context under which cryptography is used, has resulted in methods that in reality help the powerful more than the powerless. He calls on scientists to speak up when they see their technology being misused, and cites physicists' movement towards nuclear disarmament asking cryptographers to do the same.

Similarly, AI researchers should learn about the ways in which their technology is being used, question the direction institutions are moving in, and engage with other disciplines to learn from their approaches. Instead of doing parachute science, those studying fairness accountability transparency and ethics in AI should forge collaborations across disciplinary, geographic, demographic, institutional and socioeconomic boundaries, and help lift the voices of those who are marginalized. In order to work towards AI that does not further marginalize those who have historically been (and continue to be) sidelined, the educational system and general attitude amongst researchers and practitioners needs to fundamentally change and move away from the myth of meritocracy and "the view from nowhere."

---

[47] Rogaway, Phillip. "The Moral Character of Cryptographic Work." *IACR Cryptology ePrint Archive* 2015 (2015): 1162.